\documentclass[twocolumn]{aastex701}
\journalinfo{}

\usepackage{amsmath}
\usepackage{bm}
\usepackage{booktabs}
\usepackage{multirow}
\usepackage{CJKutf8}

\begin{document}

\title{SIFARI: Self-Supervised Interferometric Fitting for Astronomical Radio Imaging}

\author[orcid=0009-0006-0889-3132]{Shunyuan Mao (\begin{CJK*}{UTF8}{gbsn}毛顺元\end{CJK*})}
\affiliation{Department of Physics and Astronomy, Rice University,
6100 Main St, Houston, TX 77005, USA}
\affiliation{Rice Space Institute, Rice University,
6100 Main St, Houston, TX 77005, USA}
\affiliation{National Radio Astronomy Observatory,
520 Edgemont Road, Charlottesville, VA 22903, USA}
\affiliation{NSF-Simons AI Institute for Cosmic Origins,
Austin, TX 78712, USA}
\email[show]{shunyuanm@gmail.com}

\author[orcid=0000-0001-8061-2207]{Andrea Isella}
\affiliation{Department of Physics and Astronomy, Rice University,
6100 Main St, Houston, TX 77005, USA}
\affiliation{Rice Space Institute, Rice University,
6100 Main St, Houston, TX 77005, USA}
\email[show]{isella@rice.edu}

\author[0000-0002-2816-3229]{Paris Perdikaris}
\affiliation{Department of Mechanical Engineering and Applied Mechanics, University of Pennsylvania,
220 S 33rd St, Philadelphia, PA 19104, USA}
\email{pgp@seas.upenn.edu}

\author[orcid=0000-0001-6244-9696]{Li-Ta Lo}
\affiliation{Computing and Artificial Intelligence Division, Los Alamos National Laboratory,
Los Alamos, NM 87545, USA}
\email{ollie@lanl.gov}

\author[orcid=0000-0003-3556-6568]{Hui Li}
\affiliation{Theoretical Division, Los Alamos National Laboratory,
Los Alamos, NM 87545, USA}
\email{hli@lanl.gov}

\begin{abstract}
Radio-interferometric images are reconstructed from sparsely sampled visibilities, and CLEAN-based imaging can struggle with spatial filtering, complex morphologies, and uncertainty quantification.
Alternative methods that fit visibilities directly can address some of these limitations but often require manual choices of image priors and model hyperparameters.
We present SIFARI ({\it Self-Supervised Interferometric Fitting for Astronomical Radio Imaging}), a self-supervised neural network workflow that represents sky brightness as a continuous function of position and fits measured visibilities without an external image training set or explicit spatial regularizer.
An empirical rule sets the Fourier feature scale from the visibilities before training, controlling how readily the network fits fine structure.
Sampling network weights with {\it Stochastic Weight Averaging-Gaussian} (SWAG) gives approximate brightness uncertainty estimates, which we combine with a thermal-noise floor to construct spatially resolved signal-to-noise maps.
In synthetic ALMA tests, SIFARI yields an effective point-source response about eight times narrower than the natural-weighting CLEAN restoring beam and recovers more extended flux than CLEAN when short baselines are missing.
It also achieves higher image fidelity than the restored CLEAN images in all three morphology benchmarks.
Applied to ALMA observations of PDS~70, SIFARI recovers the bright outer ring together with faint compact emission in the central cavity.
For long-baseline-only WISPIT~2 data, SIFARI supplies a sky model for phase self-calibration where the CLEAN model is inadequate.
The restored, self-calibrated SIFARI image has approximately 30\% lower RMS noise than the CLEAN image made from the original visibilities without self-calibration.

\end{abstract}

\keywords{\uat{Astronomical methods}{1043}, \uat{Astronomy image processing}{2306}, \uat{Neural networks}{1933}, \uat{Observational astronomy}{1145}, \uat{Protoplanetary disks}{1300}, \uat{Radio interferometry}{1346}}

\section{Introduction}\label{sec:introduction}

Modern radio interferometers, including the Atacama Large Millimeter/submillimeter Array (ALMA), the Karl G. Jansky Very Large Array (VLA), MeerKAT, and the future Square Kilometre Array (SKA) and next-generation VLA (ngVLA), resolve faint and morphologically complex sources at high angular resolution.
These facilities routinely produce images with dynamic ranges greater than $1000$, but their scientific return still depends on how well sparse visibility measurements are converted into images.
Visibilities sample the Fourier transform of the sky brightness, so incomplete coverage leaves some spatial structure to be inferred during reconstruction.
CLEAN and its extensions remain the dominant tools for image reconstruction and self-calibration.

CLEAN has been successful across many applications \citep{hogbom1974aperture,cornwell2008multiscale,rau2011multi}, but several of its assumptions become restrictive for modern interferometric datasets.
CLEAN's fixed component shapes can make complex emission difficult to reconstruct.
Standard Högbom CLEAN represents emission with point sources \citep[Dirac delta functions;][]{hogbom1974aperture}, whereas Multi-Scale CLEAN uses two-dimensional Gaussians over a set of scales \citep{cornwell2008multiscale}.
These bases are effective for simple or marginally resolved emission, but they can struggle with well-resolved, non-Gaussian structures.
For complex sources, users often run many low-gain iterations inside hand or automatically defined masks, or CLEAN boxes, that restrict deconvolution to expected emission regions.
The extent of the emission is unknown before imaging, so defining these masks may require substantial user interaction and knowledge of the source.
An incomplete or inaccurate mask can limit the image dynamic range.

When short baselines are missing, CLEAN can underestimate the integrated flux and distort the recovered morphology of extended sources.
CLEAN begins from a dirty image formed by Fourier inversion of sparsely sampled visibilities.
The missing short spacings suppress large-scale emission in this image and can produce strong artifacts, including ``negative bowls'' around bright structures.

Adding the residual map to the convolved CLEAN model can bias flux measurements, especially for extended sources or shallow deconvolutions.
The restored image is the sum of the model convolved with a Gaussian restoring beam and the residual map.
The dirty beam, the point-source response set by the sampled {\it uv} coverage, can have a central lobe that differs substantially from this Gaussian beam.
The convolved model is reported in units of Jansky per CLEAN beam, whereas the residual map remains in Jansky per dirty beam.
Weighting and/or tapering the {\it uv}-data can make the dirty beam more Gaussian, but at the cost of sensitivity.
Post-deconvolution corrections such as the Jorsater \& van Moorsel (JvM) correction can rescale residuals \citep{jorsater1995high,czekala2021molecules}, but they must be applied carefully to avoid misinterpreted fluxes and underestimated noise \citep{casassus2022variable}.

CLEAN does not directly quantify uncertainty in the reconstructed flux and morphology.
CLEAN produces a single reconstructed model rather than a posterior distribution over images.
The uncertainty is commonly summarized by the RMS noise measured in an apparently empty region of the restored image.
That off-source RMS measures the thermal sensitivity away from target emission, but it does not quantify the amplitude uncertainty of bright or complex emission itself (e.g., \citealt{czekala2015disk}; \citealt{arras2019unified}).
Residual sidelobes and unmodeled source structure can also contaminate this background noise estimate.

Finally, CLEAN depends on visibilities whose amplitudes and absolute phases can be Fourier inverted.
In Very Long Baseline Interferometry and optical/infrared interferometry, atmospheric phase corruption often precludes standard phase referencing.
Imaging then relies on closure quantities, such as closure phases and amplitudes, that cancel antenna-based errors but cannot be directly Fourier-transformed into a dirty image.

Forward modeling allows reconstruction methods to fit visibilities or closure quantities by predicting these measurements from a candidate sky model.
Regularized Maximum Likelihood (RML) methods formulate imaging as an optimization problem: a likelihood term measures the discrepancy between observed and model visibilities and/or closure quantities, and spatial regularizers encode assumptions about the source morphology.
Common regularizers include image entropy \citep[which favors global continuous smoothness,][]{gull1978image}, sparsity \citep[which favors compactness,][]{Wiaux2009}, and total (squared) variation \citep[which favors constant regions separated by sharp edges,][]{Akiyama2017}.
Software libraries such as SMILI \citep[][and references therein]{akiyama2019first}, EHT-imaging \citep{chael2016high}, and MPoL \citep{czekala2021molecules} have used RML to image sources ranging from supermassive black holes to protoplanetary disks.

In pixel-based RML, each pixel intensity is a fitted parameter, so large, finely sampled images require many parameters.
A single-pointing image can sometimes be restricted to a region smaller than the primary beam, but it must still cover the astronomical source.
The grid must adequately sample the observed spatial frequencies, with pixels no larger than about one-third of the synthesized beamwidth.
Neighboring pixel intensities are correlated on the scale of the beam.
For example, at 230 GHz, ALMA's 12-meter antennas have a primary beam of approximately 22 arcseconds.
In the most extended configuration, with a maximum baseline of roughly 15 km, the angular resolution is approximately 18 mas.
Nyquist-sampling that 18 mas beam with roughly 6 mas pixels across a large fraction of the primary beam requires millions of pixels.

Fitting millions of correlated pixel intensities is an ill-posed inverse problem.
Deterministic RML methods constrain the solution with spatial regularizers such as Total Variation or Maximum Entropy \citep[see, e.g.,][]{zawadzki2023regularized, Carcamo2017}, but choosing regularizer weights often requires extensive parameter sweeps and can introduce confirmation bias \citep{akiyama2019first}.
Even among statistically acceptable fits, the recovered structure depends on the chosen regularizers and weights \citep{akiyama2019first}.
These methods return a point estimate rather than a posterior distribution over images.

Compressive sensing and Bayesian methods can also be computationally expensive on large image grids.
Compressive sensing algorithms such as SARA \citep{wiaux2009compressed,carrillo2012sparsity} impose sparsity across a fixed, redundant multi-scale wavelet dictionary, but constrained convex optimization over millions of pixels and overlapping bases is computationally expensive.
Bayesian methods such as RESOLVE \citep{junklewitz2016resolve,arras2019unified} model posterior uncertainty more directly, but sampling millions of highly correlated spatial parameters is often prohibitive for modern interferometric datasets.

Some deep-learning methods use trained neural networks to regularize or update the reconstructed image.
Hybrid Plug-and-Play (PnP) architectures, such as AIRI \citep{terris2023image}, replace analytic regularizers with trained neural denoisers within iterative imaging algorithms.
R2D2 \citep{aghabiglou2024r2d2} uses a sequence of iteration-specific neural networks to estimate residual images, analogous to learned CLEAN minor cycles.
However, for supervised methods, accuracy beyond the source and noise conditions covered by training and validation cannot be assumed.
Radio ground truth is scarce, so training often uses simulations or processed images from other wavebands that may not capture radio-source morphologies and brightness contrasts \citep{terris2023image,aghabiglou2024r2d2}.
For networks trained on simulated observations, including R2D2, this concern extends to {\it uv} coverage and instrument configuration \citep{aghabiglou2024r2d2}.

We introduce SIFARI, a self-supervised framework that uses a coordinate-based neural network to represent sky brightness as a continuous function of position.
The same network parameters determine the brightness at every sky position, so their number does not grow with the image grid used to evaluate the model.
We train the network by Fourier-transforming its predicted images and fitting the resulting model visibilities directly to the observations.
The network architecture provides an implicit image prior, and each target is reconstructed from its own measured visibilities without explicit spatial regularizers.

An overly smooth model can suppress compact emission, whereas an overly flexible model can introduce structure that the data poorly constrain.
We control this flexibility using Fourier features: sinusoidal functions of sky position supplied as network inputs.
Their frequency scale determines how readily the network fits fine structure.
We estimate this scale from the measured visibilities before training.

To assess uncertainty in the recovered emission, we use Stochastic Weight Averaging-Gaussian (SWAG) \citep{izmailov2018averaging,maddox2019simple} to approximate the posterior distribution over network weights.
Sampling this distribution produces an ensemble of sky images, whose brightness variations provide spatially resolved uncertainty estimates.

We compare SIFARI with CLEAN on synthetic ALMA benchmarks that test compact-source resolution, flux recovery under missing short spacings, and reconstruction of complex morphologies.
We apply the method to PDS~70 to examine compact emission near bright extended structure, and to WISPIT~2 to test whether the reconstructed image can serve as a self-calibration model.

\section{Method}\label{sec:method}

\begin{figure*}[tbp]
    \centering
    \includegraphics[width=0.8\textwidth]{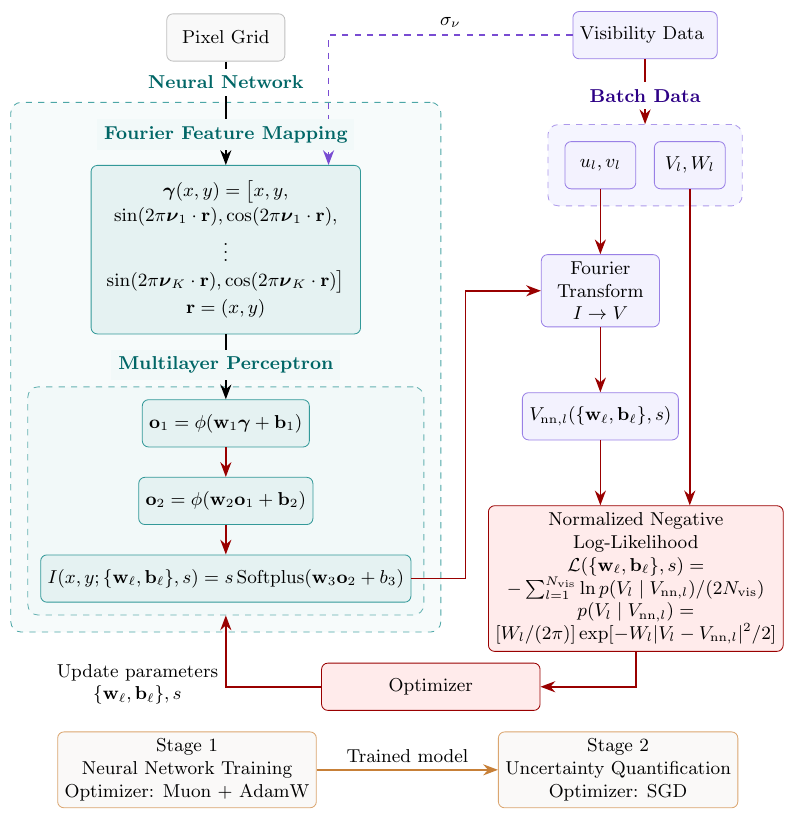}
    \caption{Two-stage SIFARI workflow for visibility-domain image reconstruction and uncertainty quantification. A coordinate-based neural network maps sky coordinates $(x, y)$ to brightness $I(x, y)$. Before training, the visibility data set the frequency scale $\sigma_\nu$ used to choose $\boldsymbol{\nu}_k$ (dashed arrow). The model image is Fourier-transformed at the observed $(u_l,v_l)$ coordinates to predict visibilities, which are compared with the data through the likelihood. The weights $W_l$ are inverse noise variances of the real and imaginary visibility components. Stage 1 uses Muon+AdamW to fit the network and yields the reported best-fitting reconstruction. Stage 2 starts from this fit, uses SGD to build the SWAG approximation, and samples network weights to estimate the pixel-wise brightness uncertainty.}
    \label{fig:training_loop}
\end{figure*}

SIFARI produces a reconstructed sky image and an uncertainty map in two stages (Figure~\ref{fig:training_loop}).
In Stage 1, we fit the neural brightness model by comparing its Fourier-transformed images with the observed visibilities, using a likelihood that accounts for measurement uncertainties.
The best-fitting Stage 1 image is the reported SIFARI image reconstruction. 
Stage 2 starts from this best-fitting model and uses Stochastic Weight Averaging--Gaussian (SWAG) to sample network weights and generate an ensemble of images.
The brightness variations across these images provide a pixel-wise uncertainty map.

\subsection{Coordinate-Based Image Model}\label{sec:architecture}

We represent sky brightness with a coordinate-based neural network, also called a neural field \citep{park2019deepsdf,mescheder2019occupancy,sitzmann2020implicit,mildenhall2021nerf,wu2022neural,levis2022gravitationally,levis2025revealing,feng2026dynamic}.
For any sky position $(x,y)$, the network predicts a brightness $I(x,y)$ using the same weights and biases across the image.
We implement this continuous brightness function as a {\it Multi-Layer Perceptron} (MLP).
Its layers form weighted sums of their inputs, add biases, and apply nonlinear activation functions.

\subsubsection{Fourier Feature Mapping}
\label{sec:fourier_feature_mapping}

An MLP trained on coordinates alone tends to learn smooth, large-scale structure more readily than fine structure, a behavior known as \textit{spectral bias} \citep{rahaman2019spectral}.
This bias makes it harder to reconstruct compact emission alongside smooth extended structure.
We use Fourier feature mapping \citep{tancik2020fourier} to supply sinusoidal functions of position alongside the coordinates.
These additional inputs provide spatial variations on scales set by their frequencies, making fine structure easier for the network to fit.

At each sky position $\mathbf{r}=(x,y)$, we concatenate the raw coordinates with $K$ sine--cosine pairs, one for each spatial frequency vector $\boldsymbol{\nu}_k$:
\begin{equation}\label{eq:fourier_mapping}
\begin{split}
    \boldsymbol{\gamma}(x, y) = \bigl[&x, y, \sin(2\pi \boldsymbol{\nu}_1 \cdot \mathbf{r}), \cos(2\pi \boldsymbol{\nu}_1 \cdot \mathbf{r}), \ldots, \\
    &\sin(2\pi \boldsymbol{\nu}_K \cdot \mathbf{r}), \cos(2\pi \boldsymbol{\nu}_K \cdot \mathbf{r})\bigr]
\end{split}
\end{equation}
We use $K=512$ frequency vectors, so the input vector $\boldsymbol{\gamma}(x,y)$ has $2+2K=1026$ components.

We choose the frequency magnitudes from a half-normal distribution, which concentrates features at low spatial frequencies while retaining a high-frequency tail.
The magnitudes $q_k=|\boldsymbol{\nu}_k|$, in units of arcsec$^{-1}$, have the probability density
\begin{equation}
    p(q) = \sqrt{\frac{2}{\pi}}\,\frac{1}{\sigma_\nu}
    \exp\left(-\frac{q^2}{2\sigma_\nu^2}\right), \qquad q \ge 0.
\end{equation}
where $\sigma_\nu$ is the scale of the distribution.
We obtain $q_k$ by treating Sobol' sequence values \citep{sobol1967distribution} as cumulative probabilities and applying the inverse cumulative distribution function (CDF).
For each magnitude, we independently draw a direction angle $\phi_k$ uniformly between 0 and $2\pi$ and set $\boldsymbol{\nu}_k = (q_k\cos\phi_k, q_k\sin\phi_k)$.
The uniform angles give the frequency distribution no preferred orientation.
The frequency vectors $\{\boldsymbol{\nu}_k\}_{k=1}^{K}$ remain fixed throughout training.

Increasing $\sigma_\nu$ gives higher frequency magnitudes and therefore more rapidly varying input features.
We estimate $\sigma_\nu$ from the radial distribution of the sampled {\it uv} coverage and the visibility signal-to-noise ratio in each baseline-length interval.
This SNR profile depends jointly on source morphology, source brightness, and the measurement noise level.
Appendix~\ref{app:auto_sigma} gives the estimator details.

\subsubsection{Network Architecture}

The MLP maps the input vector $\boldsymbol{\gamma}(x,y)$ to a brightness value through two fully connected hidden layers and a scalar output layer:
\begin{equation}
\label{eq:nn}
\begin{aligned}
    \mathbf{o}_1 &= \phi(\mathbf{w}_1 \boldsymbol{\gamma} + \mathbf{b}_1) \\
    \mathbf{o}_2 &= \phi(\mathbf{w}_2 \mathbf{o}_1 + \mathbf{b}_2) \\
    I(x, y) &= s\times\mathrm{Softplus}(\mathbf{w}_3 \mathbf{o}_2 + b_3)
\end{aligned}
\end{equation}
Here, $\mathbf{o}_1,\mathbf{o}_2\in\mathbb{R}^{128}$ are the hidden-layer outputs.
We use the Exponential Linear Unit (ELU) as the activation function $\phi$ in both hidden layers for all reconstructions.
During fitting, we adjust the weight matrices $\mathbf{w}_\ell$ and biases $\mathbf{b}_\ell$.
For the adopted architecture, the three weight matrices have dimensions $128\times1026$, $128\times128$, and $1\times128$, respectively.
The corresponding biases are 128-dimensional vectors in the hidden layers and a scalar in the output layer.
Including the brightness scale $s$, the network has 148,098 trainable parameters.

The final layer applies $\mathrm{Softplus}(z) = \ln(1 + e^z)$ to the scalar $z = \mathbf{w}_3 \mathbf{o}_2 + b_3$, producing a positive value.
The learnable parameter $s$ multiplies this output to set the overall brightness scale.
For initialization, we estimate the total flux from the mean of the 10 largest observed visibility amplitudes.
This estimate need only provide a rough starting scale, since $s$ is adjusted during fitting.

\subsection{Visibility-Domain Fitting}\label{sec:training}

To fit the brightness model to the observations, we first predict visibilities by Fourier-transforming the network's output image and define a loss function that measures their agreement with the data. We then describe how the network parameters are optimized to minimize this loss.

\subsubsection{Forward Model and Loss Function}

We evaluate $I(x,y)$ on a regular $N_\mathrm{pix}\times N_\mathrm{pix}$ grid covering the chosen field of view to form a model image.
We then use a non-uniform fast Fourier transform to compute the model visibilities $V_{\mathrm{nn},l}$ at the observed $uv$ coordinates $(u_l,v_l)$ \citep{czekala2025million}.

We model each observed visibility $V_l$ as its predicted value $V_{\mathrm{nn},l}$ plus independent, zero-mean Gaussian noise in the real and imaginary components.
Both components have variance $1/W_l$, where $W_l$ is the visibility weight shown in Figure~\ref{fig:training_loop}.
The resulting likelihood is
\begin{equation}
    p(V_l \mid V_{\mathrm{nn},l}) = \frac{W_l}{2\pi} \exp\left(-\frac{W_l |V_l-V_{\mathrm{nn},l}|^2}{2}\right).
\end{equation}

For $N_\mathrm{vis}$ independent visibility measurements, we minimize the negative log-likelihood normalized by $2N_\mathrm{vis}$:
\begin{equation}\label{eq:loss}
\begin{split}
    \mathcal{L} &= -\frac{1}{2N_\mathrm{vis}} \sum_{l=1}^{N_\mathrm{vis}} \ln p(V_l \mid V_{\mathrm{nn},l}) \\
    &= \frac{\ln(2\pi)}{2} - \frac{\sum_{l=1}^{N_\mathrm{vis}} \ln W_l}{2N_\mathrm{vis}} \\
    &\quad + \frac{\sum_{l=1}^{N_\mathrm{vis}} W_l |V_l-V_{\mathrm{nn},l}|^2}{4N_\mathrm{vis}}.
\end{split}
\end{equation}
The first two terms do not depend on the network parameters, so minimizing $\mathcal{L}$ is equivalent to minimizing the weighted squared visibility residuals.

To compare the residual amplitudes with the expected noise, we use the approximate reduced chi-squared:
\begin{equation}\label{eq:reduced_chi_squared}
    \chi_r^2 = \frac{\sum_{l=1}^{N_\mathrm{vis}} W_l |V_l-V_{\mathrm{nn},l}|^2}{2N_\mathrm{vis}}.
\end{equation}
The normalization averages the weighted squared residuals over the $2N_\mathrm{vis}$ real and imaginary components. For correctly specified noise, residuals at the expected noise level give $\chi_r^2\simeq1$.

\subsubsection{Training and Loss Minimization}

{\it Initialization:} We initialize the hidden-layer weights using Kaiming-normal initialization \citep{he2015delving}.
We initialize the overall brightness scale $s$ using the visibility-amplitude estimate described above. Before training, we sample and fix the Fourier feature frequencies $\{\nu_k\}_{k=1}^{K}$ using the visibility-based scale described in Section~\ref{sec:fourier_feature_mapping}.

{\it Iterative Fitting:} We train the model in epochs, each consisting of one pass through the full visibility dataset.
At the start of each epoch, we shuffle the full set of visibility indices once and partition the resulting permutation into disjoint mini-batches.
For each mini-batch, we compute the model visibilities and the loss (Equation~\ref{eq:loss}), then obtain the gradients by backpropagation.
Before each optimizer update, we clip the Euclidean norm of the full gradient vector to 1.0 \citep{pascanu2013difficulty}.
We train each case for 300 epochs in Stage 1.

We use {\it Muon} \citep{jordan2024muon} to update the hidden-layer weight matrices ($\mathbf{w}_1$ and $\mathbf{w}_2$), and {\it AdamW} \citep{loshchilov2017decoupled} to update the output-layer weight matrix ($\mathbf{w}_3$), all biases, and the flux scale $s$.
The initial learning rates are $2\times10^{-2}$ for {\it Muon} and $10^{-3}$ for {\it AdamW}.
We apply decoupled weight decay with a coefficient of $10^{-5}$ \citep{loshchilov2017decoupled}.

{\it Monitoring Convergence:}
At the end of each epoch, we evaluate the total loss using the entire visibility dataset.
If this loss fails to improve for several epochs, we reduce the learning rates of both optimizers so they take smaller updates.
We inspect the loss and reconstructed images across epochs to confirm that both stabilize by the end of training.
We retain the model with the lowest loss across all epochs, report its image as the Stage 1 reconstruction, and use it to initialize Stage 2 uncertainty estimation.

\subsection{Uncertainty Quantification}
\label{sec:uncertainty_quantification}
Stage 2 uses the {\it Stochastic Weight Averaging Gaussian} (SWAG) method presented in \cite{izmailov2018averaging,maddox2019simple} to assess uncertainty in the reconstructed brightness.
This method generates an ensemble of images by sampling network weights, without training multiple independent networks.

Starting from the best-fitting Stage 1 model, we continue optimization with {\it Stochastic Gradient Descent} (SGD) at a constant learning rate for 500 epochs.
After a fixed warm-up period, we save the network weights once per epoch.
SWAG uses these weight snapshots to construct a Gaussian approximation to the distribution of plausible network weights \citep{maddox2019simple}.

We draw network weights from the SWAG approximation and evaluate the corresponding images.
Their pixel-wise standard deviation, $\sigma_\mathrm{SWAG}(x,y)$, quantifies the spread in reconstructed brightness across these samples, which we use as an approximate uncertainty estimate.

We define a spatially resolved signal-to-noise ratio (SNR) diagnostic using the SWAG dispersion and a thermal-noise floor:
\begin{equation}
\label{eq:snr}
\mathrm{SNR}(x,y)=\frac{I_\mathrm{SIFARI}(x,y)}{\sqrt{\sigma_\mathrm{SWAG}(x,y)^2+\sigma_{th}^2}},
\end{equation}
where $I_\mathrm{SIFARI}(x,y)$ is the reported best-fitting Stage 1 image and $\sigma_{th}$ is the theoretical thermal noise in Jy\,arcsec$^{-2}$,
\begin{equation}
\sigma_{th} = \frac{1}{A_\mathrm{beam}\sqrt{\sum_{l=1}^{N_\mathrm{vis}} W_l}},
\end{equation}
and $A_\mathrm{beam}$ is the beam area in arcsec$^2$\,beam$^{-1}$.
The thermal-noise term sets a lower bound on the SNR denominator when $\sigma_\mathrm{SWAG}$ becomes small in low-intensity regions.

\subsection{Hyperparameter Summary}

Table~\ref{tab:static_hyperparameters} summarizes the network architecture and training settings used throughout the experiments unless stated otherwise.
The Fourier feature scale is estimated separately from each visibility dataset as described in Section~\ref{sec:fourier_feature_mapping}.
The SWAG settings are guided by \citet{maddox2019simple}.

\begin{table*}[t]
    \centering
    \caption{Default hyperparameters used in all experiments unless otherwise specified. Synthetic models and their observing setups are listed in Table~\ref{tab:case_specific_hyperparameters}.}
    \label{tab:static_hyperparameters}
    \begin{tabular}{l l c}
        \toprule
        \textbf{Category} & \textbf{Hyperparameter} & \textbf{Value} \\
        \midrule
        \multirow{5}{*}{Architecture}
            & Neurons per hidden layer $M$ & 128 \\
            & Number of hidden layers & 2 \\
            & Hidden-layer activation & ELU \\
            & Fourier feature scale $\sigma_\nu$ & Estimated from visibilities \\
            & Number of Fourier features $K$ & 512 \\
        \midrule
        \multirow{5}{*}{Stage 1: Optimization}
            & Initial Muon learning rate & $2\times10^{-2}$ \\
            & Initial AdamW learning rate & $10^{-3}$ \\
            & Weight decay & $10^{-5}$ \\
            & Gradient clipping & 1.0 \\
            & Number of epochs $N_\mathrm{epoch}$ & 300 \\
        \midrule
        \multirow{6}{*}{\shortstack[l]{Stage 2: Uncertainty\\(SWA/SWAG)}}
            & SGD learning rate $\eta_\mathrm{SWA}$ & $10^{-2}$ \\
            & SGD momentum & 0.9 \\
            & Training epochs $N_\mathrm{SWA}$ & 500 \\
            & Epoch at which snapshot collection begins & 200 \\
            & Rank of the low-rank covariance component $R$ & 20 \\
            & Number of SWAG weight samples $S$ & 30 \\
        \bottomrule
    \end{tabular}
\end{table*}

\section{Synthetic Benchmark Models}\label{sec:assessment}

We compare SIFARI with CLEAN to evaluate its performance in reconstructing and resolving compact sources (\S\ref{sec:psf} and \S\ref{sec:angular_resolution}), recovering extended emission (\S\ref{sec:spatial_filtering_flux_recovery}), and imaging complex morphologies (\S\ref{sec:image_fidelity}).
We also examine the image uncertainty maps for the complex morphology models (\S\ref{sec:uncertainty_assessment}).

We generate synthetic ALMA datasets with the CASA task \texttt{simobserve}, including thermal noise.
All simulations use a central observing frequency of $340\,\mathrm{GHz}$, a total bandwidth of $8\,\mathrm{GHz}$, and an integration time of $6\,\mathrm{s}$, with the targets positioned at zenith.
Table~\ref{tab:case_specific_hyperparameters} summarizes the ALMA configurations, on-source times, and ground-truth grid dimensions and pixel scales.
The tests that combine the C-3, C-5, and C-9 configurations have a total on-source time of 110 minutes.

\begin{table*}[t]
    \centering
    \caption{Synthetic benchmark models and their ALMA \texttt{simobserve} setups. The grid columns specify the ground-truth images used to generate the synthetic visibilities; reconstruction grids are specified separately in the text. The C-3, C-5, and C-9 configurations span baseline ranges of 15--500\,m, 15--1398\,m, and 368--13\,894\,m, respectively. Rows with multiple configurations use their combined visibilities.}
    \label{tab:case_specific_hyperparameters}
    \begin{tabular}{lccc}
        \toprule
        Source Model & ALMA configurations used & \shortstack{Ground-truth\\pixels per side} & \shortstack{Ground-truth\\pixel size (\arcsec)} \\
        \midrule
        One point source & C-3 (20\,min) + C-5 (30\,min) + C-9 (60\,min) & 1024 & 0.0005 \\
        \midrule
        Two point sources ($\Delta \mathrm{RA} = 0.009''$) & \multirow{2}{*}{C-3 (20\,min) + C-5 (30\,min) + C-9 (60\,min)} & \multirow{2}{*}{1024} & \multirow{2}{*}{0.0005} \\
        Two point sources ($\Delta \mathrm{RA} = 0.010''$) &  &  &  \\
        \midrule
        Uniform disk ($\theta_\mathrm{UD}=0.2\arcsec$) & \multirow{3}{*}{C-9 (60\,min)} & 200 & 0.0015 \\
        Uniform disk ($\theta_\mathrm{UD}=0.6\arcsec$) &  & 680 & 0.0015 \\
        Uniform disk ($\theta_\mathrm{UD}=1.0\arcsec$) &  & 960 & 0.0015 \\
        \midrule
        Inclined ring & \multirow{3}{*}{C-3 (20\,min) + C-5 (30\,min) + C-9 (60\,min)} & 1160 & 0.0015 \\
        Multiple inclined rings &  & 1580 & 0.0015 \\
        Ophiuchus (L1688) irregular cloud &  & 1380 & 0.0015 \\
        \bottomrule

    \end{tabular}
\end{table*}

\subsection{Effective Point-Source Response}\label{sec:psf}
Unlike a CLEAN model built from Dirac $\delta$ components, SIFARI represents a point source with a finite-width brightness profile.
To determine whether this representation broadens compact emission beyond the nominal angular scale of the observations, we reconstruct a point source at the phase center and measure its angular extent and recovered flux.
The injected point source has a flux of 10 mJy, and the simulated ALMA observations cover a baseline range of 15--13\,894\,m, for which $\lambda/B_{max} = 0.0131''$.
The CLEAN and SIFARI reconstructions use $1024\times1024$ grids with a pixel size of $0.0005\arcsec$.

For comparison, we reconstruct the same dataset with CLEAN using natural, Briggs robust=0.5, and uniform weighting.
The CLEAN restoring beam FWHMs are \(0.034'' \times 0.031''\) for natural weighting, \(0.025'' \times 0.022''\) for Briggs robust=0.5, and \(0.015'' \times 0.012''\) for uniform weighting.
The RMS noise in the CLEAN images is 0.0135 mJy/beam for natural weighting, 0.0153 mJy/beam for Briggs robust=0.5, and 0.0215 mJy/beam for uniform weighting.
  
Fitting the SIFARI reconstructed map with a 2D Gaussian yields an effective point-source response with a FWHM of \(0.0040'' \times 0.0037''\), about 4 times narrower than the uniform-weighted CLEAN restoring beam and 8 times narrower than the natural-weighted beam.
The recovered flux is 10.1 mJy (\(1.01\times\) the ground truth).
Thus, in this test, the finite point-source response remains well below \(\lambda/B_{\max}=0.0131''\), while the integrated flux is recovered to within 1\%.

\begin{figure}[t]
    \centering
    \includegraphics[width=1\columnwidth]{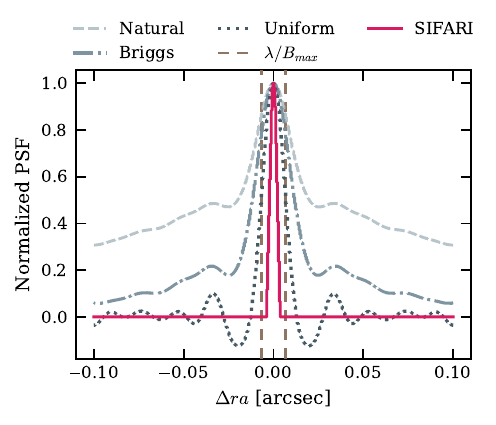}
    \caption{Normalized 1D slices along RA of the dirty beams (gray) and the SIFARI point-source response (red). Natural, Briggs, and uniform weighting are shown with dashed, dash-dotted, and dotted curves, respectively; SIFARI is shown with a solid curve. The Briggs dirty beam uses a robust parameter of 0.5. Each profile is normalized to its peak. The vertical dashed lines mark $\pm\lambda/B_{max}$.}
    \label{fig:points1_compare_clean_nn_1d_linear}
\end{figure}

The compact SIFARI response also shows no visible sidelobes in the RA slice in Figure~\ref{fig:points1_compare_clean_nn_1d_linear}.
For comparison, the uniform-weighted dirty beam has sidelobes at the 10\% level, while the natural- and Briggs-weighted dirty beams have extended shoulders around their compact cores.

\subsection{Angular Resolution}\label{sec:angular_resolution}
We next test whether the narrow response to a single point source allows SIFARI to separate two nearby sources.
We generate two synthetic models, each containing a pair of 10 mJy point sources separated by either $0.009''$ or $0.010''$.
We use the same observing setup and reconstruction grid as in \S\ref{sec:psf}.
For this test, we define the angular resolution as the source separation at which the midpoint intensity falls to 50\% of the peak intensity.

\begin{figure*}[t]
    \centering
    \includegraphics[width=0.48\textwidth]{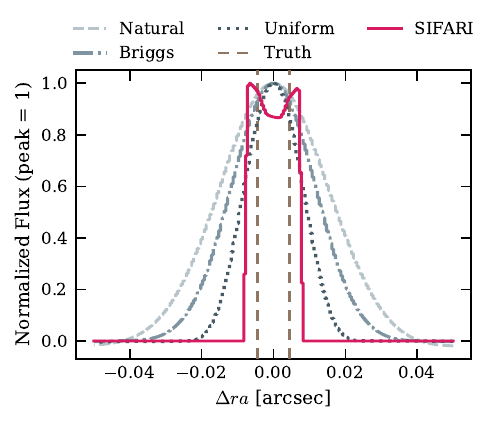}\hfill
    \includegraphics[width=0.48\textwidth]{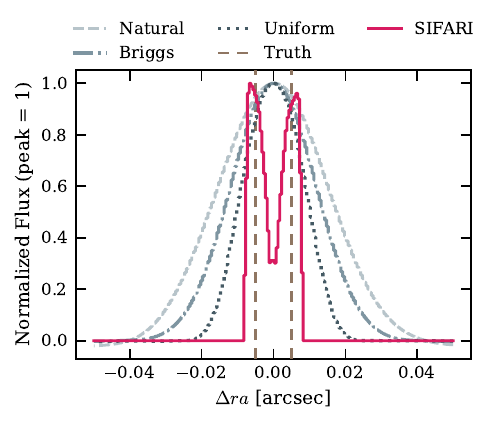}
    \caption{Two-point-source tests: normalized 1D flux slices of CLEAN and SIFARI reconstructions along the RA axis connecting the sources (linear scale). Natural-, Briggs-, and uniform-weighted CLEAN images are shown with gray dashed, dash-dotted, and dotted curves, respectively; SIFARI is shown with a red solid curve. Each profile is normalized to its peak. The vertical dashed lines mark the injected source positions. Left: \(\Delta \mathrm{RA}=0.009''\). Right: \(\Delta \mathrm{RA}=0.010''\).}
    \label{fig:points2_compare_clean_nn_1d_linear}
\end{figure*}

In these tests, the natural-, Briggs-, and uniform-weighted CLEAN images blend each pair into a single broad peak.
SIFARI, however, recovers two distinct peaks in both cases (Figure~\ref{fig:points2_compare_clean_nn_1d_linear}).
At a separation of $0.009''$, the midpoint intensity is roughly 90\% of the peak intensity, whereas at $0.010''$ it falls to about 30\%.
Under the 50\% criterion, the angular resolution for this observational configuration lies between $0.009''$ and $0.010''$, below the $\lambda/B_{max}$ scale of $0.013''$.
Guided by this angular-resolution estimate and the point-source response measured in \S\ref{sec:psf}, we adopt a reconstruction pixel size of $0.003\arcsec$ for SIFARI in the subsequent benchmarks.
These grids retain the fields of view of the corresponding ground-truth images in Table~\ref{tab:case_specific_hyperparameters}.

\subsection{Flux Recovery under Spatial Filtering}
\label{sec:spatial_filtering_flux_recovery}

We next test whether SIFARI can recover extended emission when short baselines are missing.
We generate synthetic ALMA data using only the C-9 configuration for uniform disks with angular diameters of $\theta_{UD}=0.2\arcsec$, $0.6\arcsec$, and $1.0\arcsec$, each with a total flux of 200 mJy.
We reconstruct the CLEAN images using natural weighting.
For the CLEAN and SIFARI reconstructions, we measure the disk flux within a circular aperture matching the diameter of the corresponding disk model.

Following the empirical ALMA Technical Handbook definition, the {\it Maximum Recoverable Scale} (MRS) of C-9 at $\lambda=0.88$ mm is $\theta_{\mathrm{MRS}}\simeq0.983\lambda/L_5\simeq0.24\arcsec$, where $L_5$ is the fifth percentile of the baseline-length distribution \citep{cortes_2026_18793803}.
This MRS provides a reference scale for spatial filtering rather than a sharp boundary for flux recovery.

For $\theta_{UD}=0.2\arcsec < \theta_{MRS}$, CLEAN and SIFARI recover the full 200 mJy (Table~\ref{tab:flux_comparison}).
For the $\theta_{UD}=0.6\arcsec$ and $1.0\arcsec$ disks, missing short baselines leave more of the large-scale emission unconstrained.
CLEAN recovers only 20.5\% and 10\% of the injected flux, respectively, while SIFARI recovers 101\% and 109.5\%.

\begin{figure*}[t]
    \centering
    \includegraphics[height=0.275\textwidth]{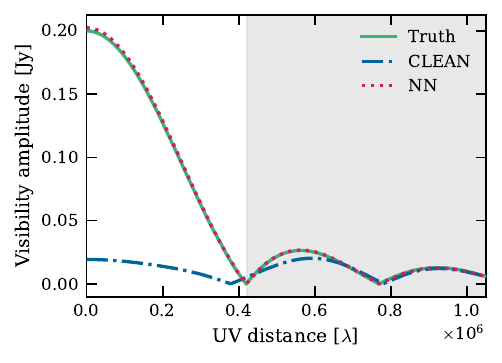}\hspace{0.015\textwidth}
    \includegraphics[height=0.275\textwidth]{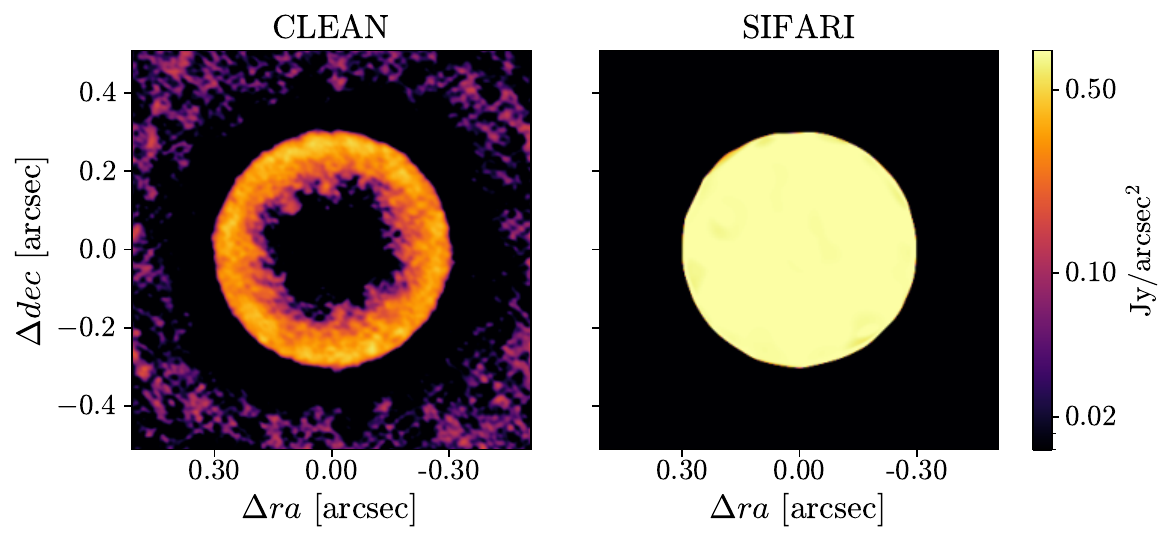}
    \caption{Visibility profiles and reconstructed images for the uniform disk with an angular diameter of $0.6\arcsec$ and total flux of 0.2 Jy. {\it Left:} Visibility amplitudes of the uniform-disk model (green solid curve), CLEAN model (blue dash-dotted curve), and SIFARI model (red dotted curve). The light-gray region marks the baseline range sampled by the observations. {\it Middle and Right:} CLEAN and SIFARI reconstructed images, respectively, shown on a common logarithmic intensity scale.}
    \label{fig:disk_r0.3_compare_clean_nn}
\end{figure*}

The CLEAN flux deficit is evident in the short-baseline visibility profile of the $0.6\arcsec$ disk (Figure~\ref{fig:disk_r0.3_compare_clean_nn}).
We calculate the model visibility amplitudes by Fourier transforming the CLEAN and SIFARI image models.
The visibility at zero baseline gives the total flux of each image model.
Within the sampled {\it uv} range (light-gray shaded region), the two profiles are broadly consistent, although SIFARI follows the uniform-disk model more closely.
At short baselines not covered by the data, the CLEAN amplitudes fall well below those of the uniform-disk model, while the SIFARI amplitudes remain close to the true values.
SIFARI extrapolates into this unsampled range using the implicit prior encoded in its positive, Fourier-feature-based representation of the sky brightness.

The impact of spatial filtering is clearly visible in the CLEAN image: only the disk edges (high spatial frequencies) are recovered, and the intensity at the disk center is suppressed to nearly zero. 
The CLEAN image also contains strong spurious emission outside the disk.
In contrast, the SIFARI image retains a nearly uniform brightness across the disk interior, with no prominent artifacts outside the disk.

\begin{table}[t]
    \centering
    \caption{Flux recovery for uniform-disk models observed with the ALMA C-9 configuration. Recovered fluxes are measured within a circular aperture with the same diameter as the corresponding disk. The CLEAN images use natural weighting.}
    \begin{tabular}{lcccc}
        \toprule
        Disk diameter    & Model Flux & CLEAN & SIFARI \\
        (\arcsec) & (mJy) & (mJy) & (mJy) \\
        \midrule
        0.2 & 200 &  200 & 200 \\
        0.6 & 200 &  41  & 202  \\
        1.0 & 200 &  20  & 219  \\
        \bottomrule
    \end{tabular}
    \label{tab:flux_comparison}
\end{table}

\subsection{Image Fidelity}
\label{sec:image_fidelity}
Our final test evaluates how well SIFARI can reconstruct images as their morphological complexity increases.  
To do this, we design three models: an inclined ring with an additional faint point source, an inclined system composed of multiple rings, and an extended, irregular structure resembling a molecular cloud.

The first model is tailored to mimic 0.8~mm ALMA observations of the PDS~70 disk, including a point source at the position of PDS~70~c \citep{isella2019detection, benisty2021circumplanetary}.
The ring has a diameter of 1.2\arcsec, a radial Gaussian profile with a width ($\sigma$) of 0.05\arcsec, and a total flux of 175~mJy.  
As in PDS~70, the ring is viewed at an inclination of $51.7^\circ$ and a position angle of $160.4^\circ$ measured East of North.  
The point source has a flux of 0.1115~mJy and is located at $(\Delta\mathrm{RA},\Delta\mathrm{Dec}) = (-0.2151\arcsec,\,0.0378\arcsec)$.  

The multiple-ring test model is inspired by ALMA DSHARP observations of HD~163296 \citep{isella2018disk}. It is represented as five concentric Gaussian rings with diameters ranging from 0.076\arcsec\ to 2.004\arcsec\ and peak surface brightnesses between 0.20 and 1.7~Jy~arcsec$^{-2}$, all sharing an inclination of $46.7^\circ$ and a position angle of $133.3^\circ$.
The total flux in the model is 0.57~Jy. 

The third model is derived from the L1688 column-density map of the Ophiuchus molecular complex \citep{ladjelate2020herschel}, obtained within the Herschel Gould Belt Survey \citep{andre2010filamentary}.
We restrict ourselves to the central region of L1688, resample it to a ground-truth pixel scale of $0.0015\arcsec$, and renormalize the intensities to a total flux of 0.1~Jy.

For each of these models, we produce synthetic datasets using the ALMA configurations summarized in Table~\ref{tab:case_specific_hyperparameters}, and then reconstruct the images with multiscale CLEAN (Briggs weighting with robust=0.5) and SIFARI.
We show both the CLEAN model and the restored CLEAN image. The latter is formed by convolving the CLEAN model with the restoring beam and adding the residual image.
For pixel-wise comparisons, we evaluate the fitted SIFARI model at the ground-truth coordinates and align the CLEAN images to the same grid, independently of the grid used during SIFARI training.

We assess reconstruction quality using three metrics that compare pixel intensities, local structure, and total flux.
The first is the classical {\it Image Fidelity Score} (IFS), defined as the mean per-pixel signal-to-error ratio between the reconstructed image $\hat{I}$ and the ground truth $I_{\text{true}}$,
\begin{equation}
IFS = \frac{1}{N_\mathrm{gt}^2}\sum_{i,j=1}^{N_\mathrm{gt}} \frac{\hat{I}(x_i, y_j)}{|\hat{I}(x_i, y_j) - I_{\text{true}}(x_i, y_j)| + \epsilon},
\end{equation}
where $N_\mathrm{gt}$ is the number of pixels per side in the ground-truth image and $\epsilon = 10^{-7}$ guards against division by zero on near-identical pixels.
The numerator is signed so that negative sidelobes in the reconstruction pull the IFS downward.  
Higher values indicate better fidelity. We average over all pixels.

To complement the pixel-wise intensity comparison provided by IFS, we use the {\it Structural Similarity Index Measure} (SSIM), as defined in Section III.B of \cite{wang2004image}. SSIM compares local luminance, contrast, and structure.
To avoid letting a small number of very bright pixels determine the SSIM dynamic range, we clip the reconstructed and ground-truth images to the range from zero to the 99th percentile of the ground-truth intensity before calculating SSIM.
An SSIM of 1 indicates a perfect match between the clipped images.
The third metric is the recovered total-flux ratio, defined as the total flux in the reconstructed image divided by the total flux in the ground-truth model.

\begin{figure*}[t!]
    \centering
    \begin{minipage}{0.95\textwidth}
        \centering
        \includegraphics[width=\textwidth]{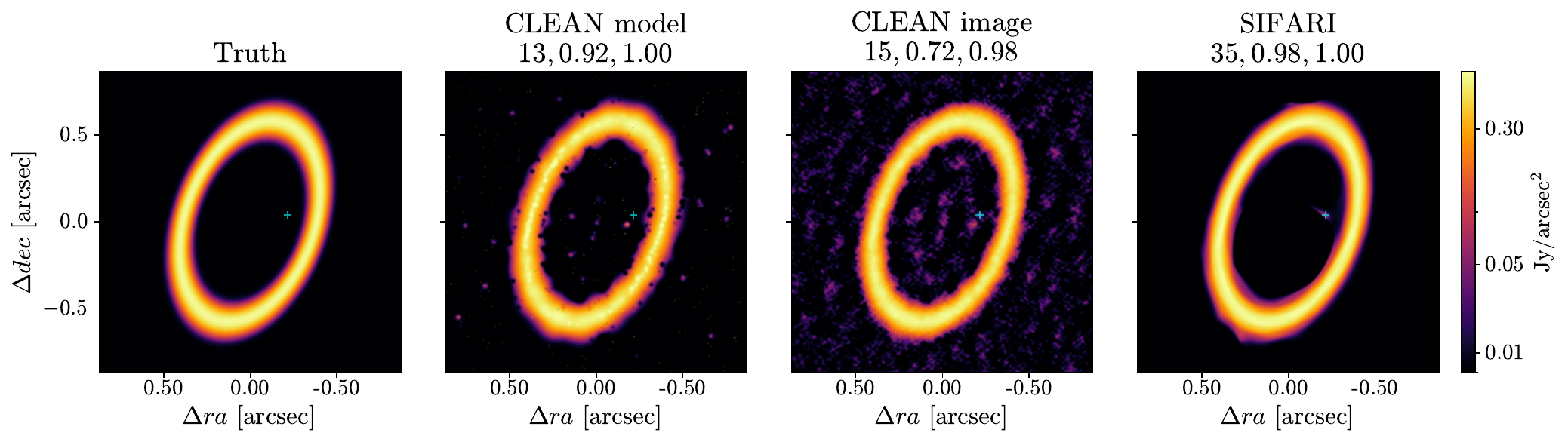}
        \\
        \includegraphics[width=\textwidth]{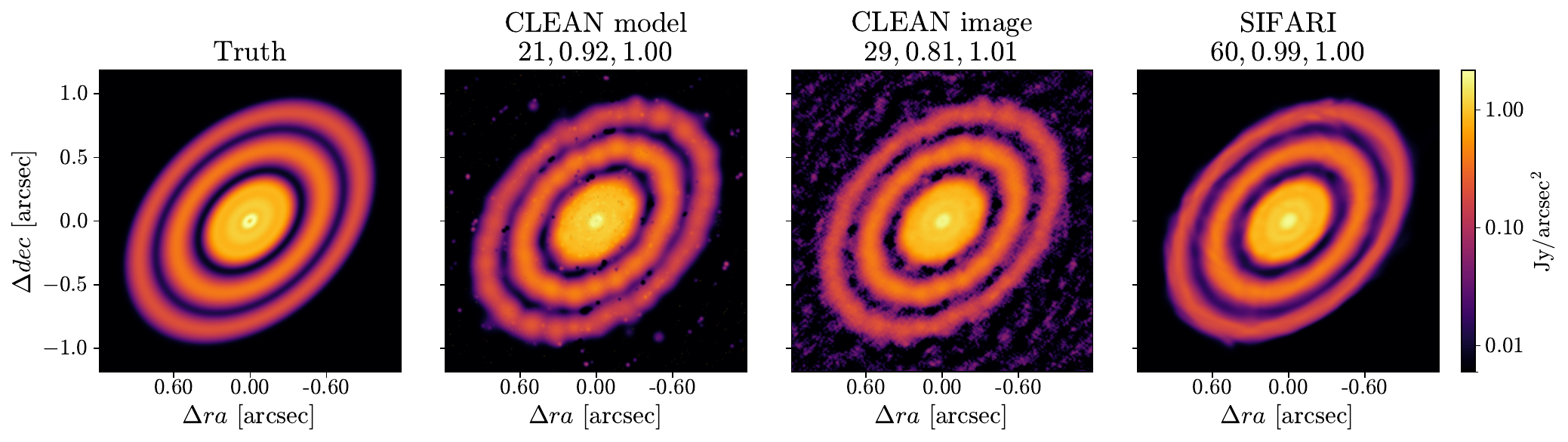}
        \\
        \includegraphics[width=\textwidth]{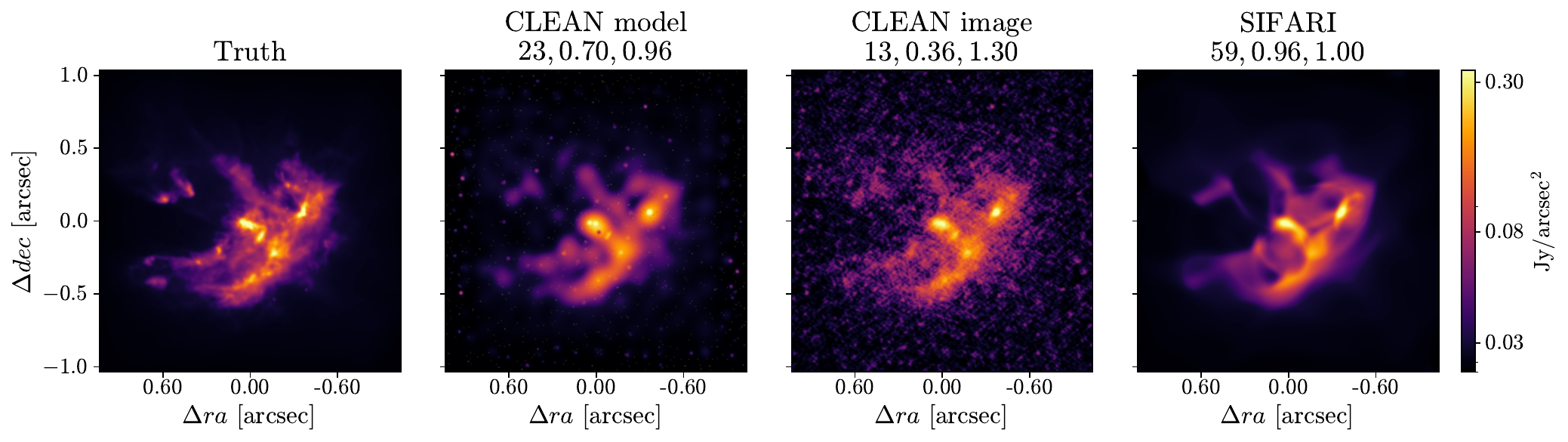}
    \end{minipage}
    \caption{
        CLEAN and SIFARI reconstructions of the three models described in Sec.~\ref{sec:image_fidelity}. Rows show the inclined ring with a point source, the multiple-ring system, and the molecular cloud, from top to bottom. Columns show the ground truth, CLEAN model, restored CLEAN image, and SIFARI reconstruction, from left to right. Each row uses a common logarithmic intensity scale. In the inclined-ring model (top row), the cyan plus sign marks the injected point-source position.
        The three numbers above each reconstruction give IFS, SSIM, and the recovered total-flux ratio, respectively.
        The FWHM of the CLEAN beam is \(0.032'' \times 0.027''\).
    }
    \label{fig:compare_clean_nn_test}
\end{figure*}
Across all three morphology benchmarks, SIFARI reconstructions achieve higher IFS and SSIM values than the restored CLEAN images, with recovered total flux ratios closer to unity (Figure~\ref{fig:compare_clean_nn_test}).
For all synthetic benchmarks in Section~\ref{sec:assessment}, the best-fitting SIFARI reconstructions yield $\chi_r^2$ values within $10^{-3}$ of unity, indicating that the overall weighted visibility-residual power is close to the level expected from the adopted noise variances.

In the inclined-ring case, the faint point source is apparent against the low background in the SIFARI reconstruction, while it is marginally detected in the restored CLEAN image.
In particular, the intensity at the point-source position in the CLEAN image is 0.06 mJy/beam, which is about half of the injected point source flux, and the rms noise is 0.02 mJy/beam, corresponding to an SNR of 3. 
The CLEAN model and restored CLEAN image also show point-like artifacts within the ring with fluxes exceeding that of the injected point source, but these features do not coincide with its position.
SIFARI recovers the point source at the correct position with a total flux of 0.08 mJy (72\% of the injected flux).

The greatest improvement in terms of image fidelity and flux recovery occurs in the molecular cloud model, where the restored CLEAN image is affected by noise, deconvolution residuals, and beam smearing, whereas the SIFARI image better preserves the emission patterns and recovers the correct flux.
In this case, the CLEAN model provides better image reconstruction than the restored CLEAN image across all three metrics.
In particular, the total flux in the restored CLEAN image is approximately 30\% higher than that of the ground-truth model, whereas the CLEAN model contains 96\% of the total flux.
The flux excess arises when the substantial diffuse, faint emission in the CLEAN residual map is added to the convolved CLEAN model.
As discussed in the Introduction, the residual image has units of Jy per {\it dirty} beam, while the convolved CLEAN model has units of Jy per {\it clean} beam.
Because the Robust=0.5 dirty beam is poorly reproduced by the Gaussian clean beam, this unit mismatch results in overestimating the residual flux and, in turn, the total flux.

\subsection{Uncertainty Maps}
\label{sec:uncertainty_assessment}

\begin{figure*}[!t]
    \centering
    \begin{minipage}{0.95\textwidth}
        \centering
        \includegraphics[width=\textwidth]{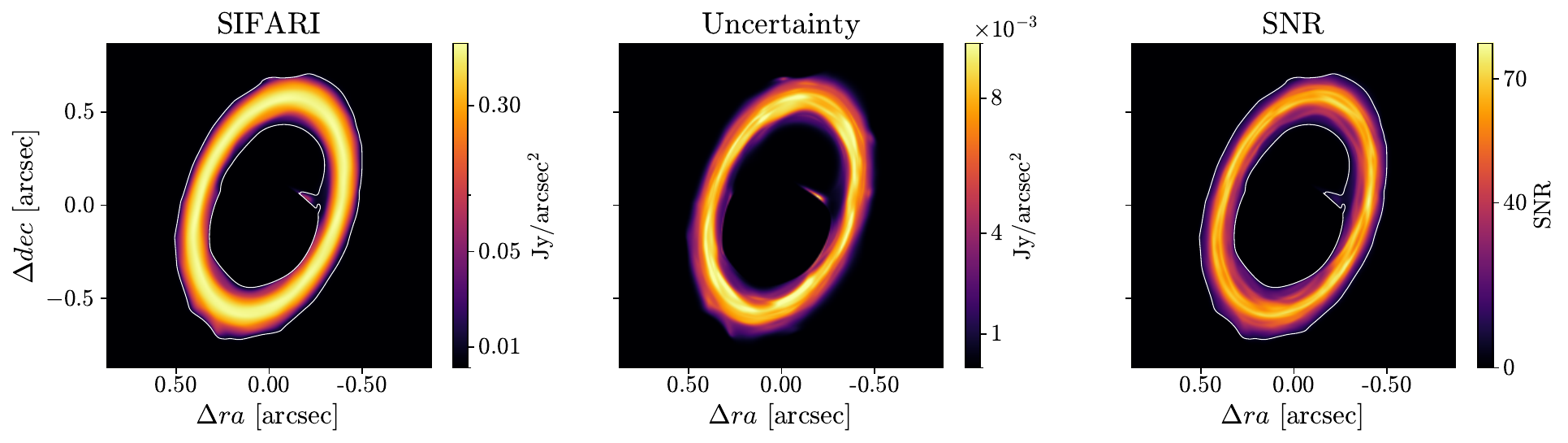}
        \\
        \includegraphics[width=\textwidth]{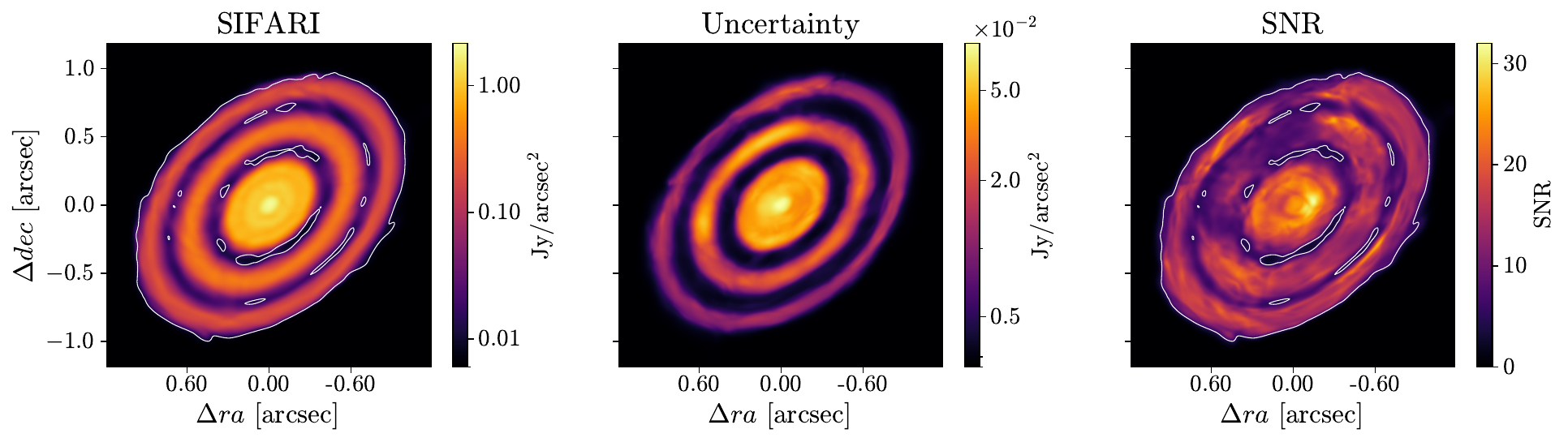}
        \\
        \includegraphics[width=\textwidth]{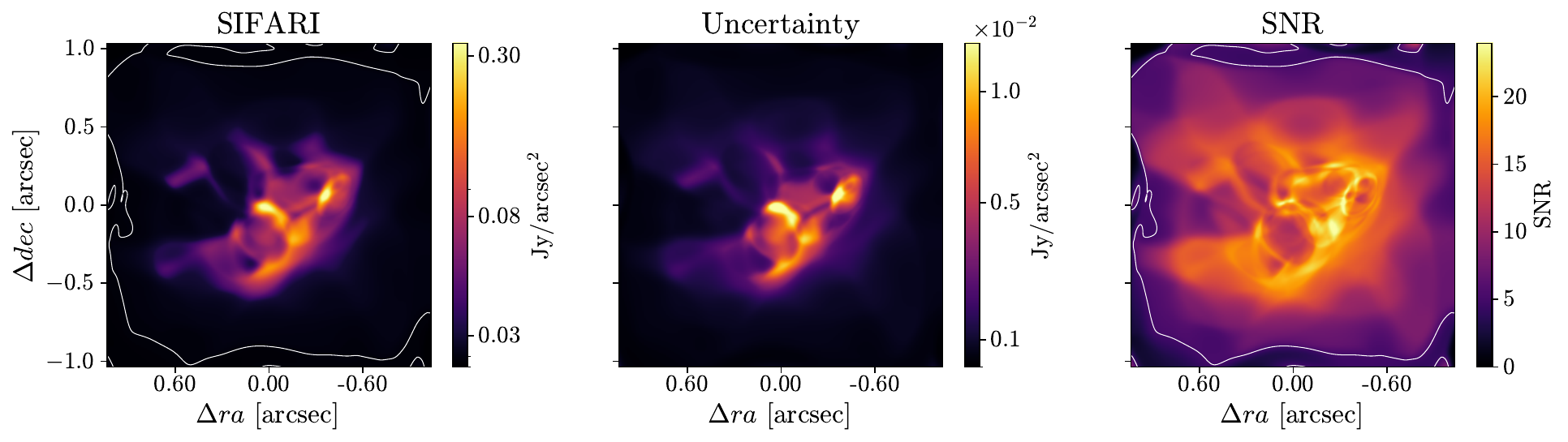}
    \end{minipage}
    \caption{
        SIFARI reconstructions and uncertainty diagnostics for the three morphology benchmarks in Figure~\ref{fig:compare_clean_nn_test}.
        Rows show the inclined ring with a point source, the multiple-ring system, and the molecular cloud, from top to bottom.
        Columns show the SIFARI reconstruction, the pixel-wise standard deviation across SWAG samples ($\sigma_\mathrm{SWAG}$), and the SNR defined by Eq.~\ref{eq:snr}, from left to right.
        The white contours on the reconstruction and SNR panels mark SNR=5.
    }
    \label{fig:compare_uq}
\end{figure*}

Having compared the reconstructions with the ground-truth images, we next examine how much the reconstructed brightness varies across SWAG samples.
Figure~\ref{fig:compare_uq} shows the SIFARI reconstructions, their associated uncertainty maps, and the SNR maps defined by Eq.~\ref{eq:snr} for the three morphology benchmarks.

In these examples, the absolute SWAG uncertainty is generally larger in brighter emitting regions and approaches zero in the off-source background.
In these off-source regions, the SNR denominator is therefore dominated by the thermal-noise term.

The peak SIFARI SNR values are approximately 32 and 28 for the multi-ring and molecular-cloud models, respectively.
In the inclined-ring benchmark, the injected point source reaches a local peak SNR of approximately 12, compared with typical values of 60--70 along the bright ring.
The SWAG term dominates the SNR denominator in both regions, with the lower point-source SNR reflecting its larger uncertainty relative to the reconstructed intensity.
We use these spatially resolved SNR maps in Section~\ref{sec:application_pds70} to examine compact emission near the bright ring in PDS~70.

\section{Application to Real Data}
\label{sec:application}

We apply SIFARI to two ALMA datasets that test different aspects of the reconstruction.
PDS~70 tests the recovery of faint, compact emission near a bright dust ring.
The long-baseline WISPIT~2 data test whether a SIFARI reconstruction can provide a self-calibration model under strong spatial filtering.

\subsection{Protoplanetary Disk of PDS 70}
\label{sec:application_pds70}

\begin{figure*}[!t]
    \centering
    \begin{minipage}{0.95\textwidth}
        \centering
        \includegraphics[width=\textwidth]{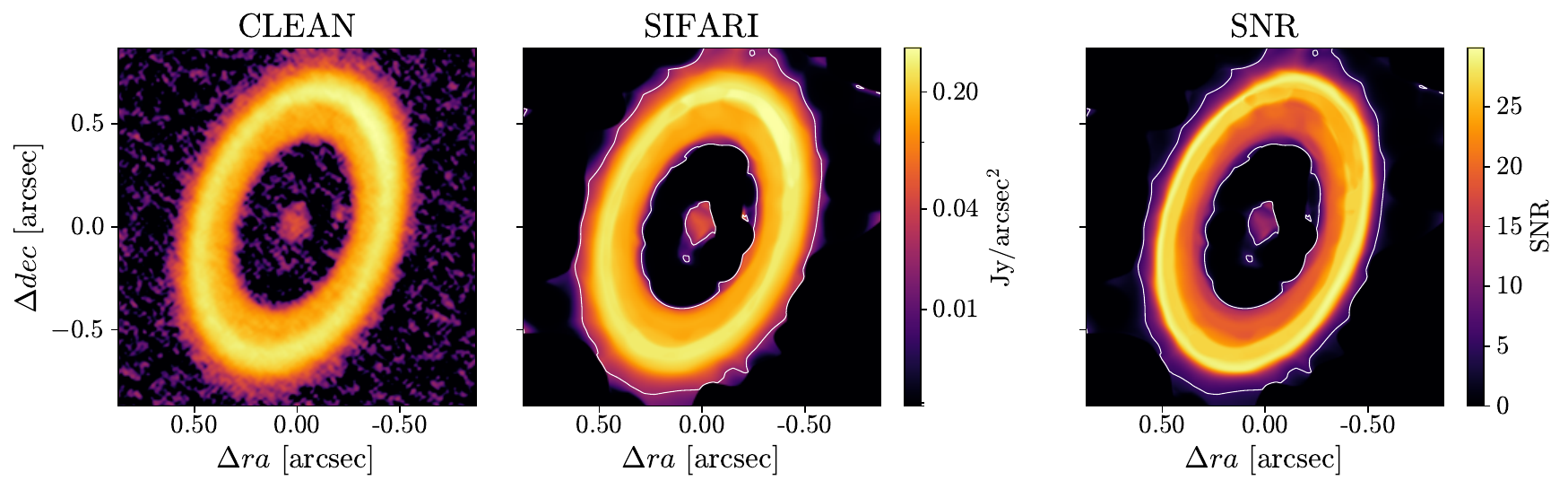}
        \vspace{0.5em}
        \includegraphics[width=\textwidth]{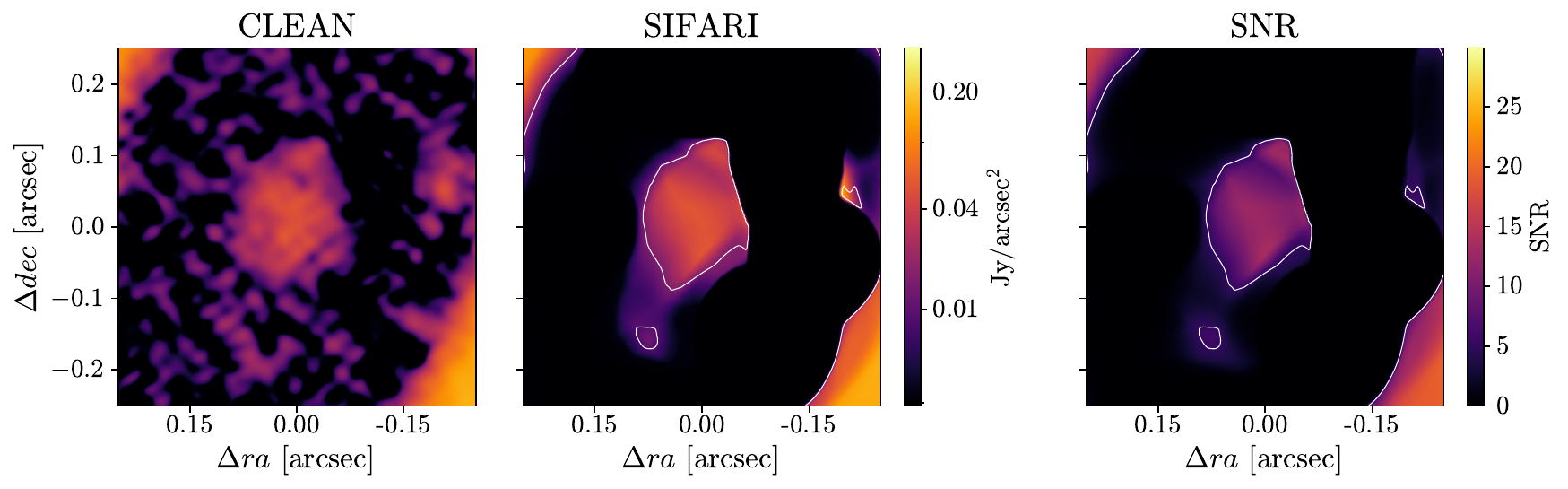}
    \end{minipage}
    \caption{Comparison of reconstructed images and signal-to-noise ratio (SNR) diagnostics for real PDS~70 data. Top: global view. Bottom: zoomed-in view of the central region, including the central emission and the locations of PDS~70~b and PDS~70~c. In each row, the panels show, from left to right, the restored CLEAN image, the SIFARI reconstruction, and the SIFARI SNR map. The CLEAN and SIFARI images share a common intensity scale. The white contours on the SIFARI and SNR panels mark SNR=5, as defined in Eq.~\ref{eq:snr}. The CLEAN beam is \(0.045'' \times 0.034''\).}
    \label{fig:pds70_real_compare_clean_nn_combined}
\end{figure*}

We apply CLEAN and SIFARI to the combined long-, intermediate-, and short-baseline ALMA continuum dataset (LB19+IB17+SB16) of PDS~70 presented by \citet{benisty2021circumplanetary}.
The field contains a bright outer dust ring, weak inner-disk emission, and candidate circumplanetary emission at the locations of PDS~70~b and PDS~70~c.
Recovering the faint compact emission requires separating it from the nearby bright ring.

We produce the CLEAN reconstruction using multiscale deconvolution with scales of 0, 1, 3, and 6 times the CLEAN beam and Briggs weighting with a robust parameter of 0.5.
We stop CLEAN at 0.03~mJy~beam$^{-1}$, approximately three times the expected thermal RMS of 0.0097~mJy~beam$^{-1}$.
The CLEAN image shows pronounced background fluctuations inside the ring, particularly near PDS~70~b, making it difficult to distinguish faint compact emission there from the background (Figure~\ref{fig:pds70_real_compare_clean_nn_combined}, lower left panel).
The local CLEAN RMS within the ring is 0.013~mJy~beam$^{-1}$, modestly above this thermal-noise estimate.
At PDS~70~c, CLEAN recovers a peak of 0.086~mJy~beam$^{-1}$, corresponding to a local peak-to-RMS ratio of 6.6, whereas no peak at PDS~70~b exceeds three times the local RMS.

The CLEAN and SIFARI reconstructions recover the same total flux of 0.177~Jy.
The SIFARI reconstruction has a visibly cleaner background inside the ring, making the compact emission at PDS~70~b easier to distinguish.
The emission at PDS~70~c is also more sharply defined than in the CLEAN image.

We then assess the reconstructed compact emission with the SWAG uncertainty and SNR maps defined in Section~\ref{sec:uncertainty_quantification}.
The SIFARI SNR uses the spatially varying SWAG dispersion and the thermal-noise floor, whereas the CLEAN peak-to-RMS ratios above use the local background RMS.
The SIFARI SNR peaks at 6.0 and 5.9 at PDS~70~c and PDS~70~b, respectively, while the bright outer ring has an average SNR of 23.
This application demonstrates that SIFARI can recover faint compact emission near a bright ring while providing spatially resolved estimates of reconstruction uncertainty.

\subsection{Self-calibration of WISPIT~2}
\label{sec:wispit2}

\begin{figure*}[!t]
    \centering
    \includegraphics[width=1\textwidth]{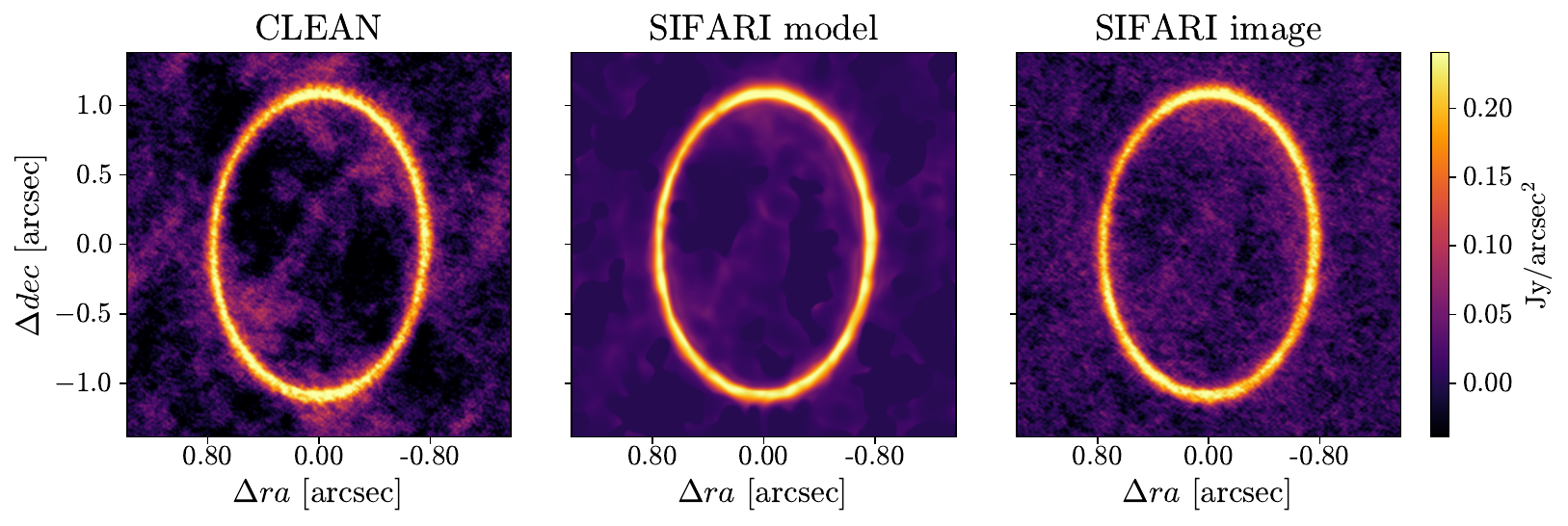}
    \caption{Three image reconstructions of the WISPIT~2 continuum, shown with a common color scale. \emph{Left:} naturally weighted CLEAN image from the original calibrated visibilities, before self-calibration. \emph{Middle:} SIFARI model fitted to the self-calibrated visibilities, without beam convolution or added residuals. \emph{Right:} restored SIFARI image, formed by convolving the middle-panel model with the restoring beam and adding the residual image from the self-calibrated visibilities.}
    \label{fig:wispit2_compare_clean_nn}
\end{figure*}

We apply SIFARI to the ALMA Band~7 ($\lambda=0.88$ mm) continuum observations of the planet-hosting system WISPIT~2 \citep{van2025wide,facchini20262}.
These observations reveal a narrow dust ring at a deprojected radius of 144.4 au, with no detected dust interior to it.
The C-10 configuration provides baselines from 132 m to 15.2 km and an angular resolution of $25\times17$ mas, but the scarcity of short baselines limits the maximum recoverable scale to $0.27\arcsec$, or roughly 36 au \citep{facchini20262}.
This scale is much smaller than the ring, leaving its large-scale emission poorly constrained.

Phase-only self-calibration uses a reconstructed sky model as a reference for correcting residual phase errors in the data.
For WISPIT~2, the CLEAN reconstruction does not provide a suitable model for this step.
Motivated by SIFARI's recovery of extended emission in the synthetic tests (Section~\ref{sec:spatial_filtering_flux_recovery}), we test whether its reconstruction can supply the model needed for self-calibration of these spatially filtered data.

We first create a CLEAN image using CASA \texttt{tclean} on the original, non-self-calibrated visibilities (Figure~\ref{fig:wispit2_compare_clean_nn}, left panel).
This image is broadly consistent with that in \citet{facchini20262}, except that we use natural weighting, whereas they applied Briggs weighting with robust = 1.0.
The resulting image exhibits substantial negative emission caused by spatial filtering and contains a total flux of 140 mJy calculated over an elliptical area with a major axis of 3\arcsec, a minor axis of 2\arcsec, and $\textrm{PA}=0\arcdeg$.

We then train SIFARI on the original visibilities and use the resulting sky model to solve for antenna phase corrections with CASA \texttt{gaincal}.
After applying these corrections to the data, we train a new network from scratch on the self-calibrated visibilities to obtain a self-calibrated SIFARI model (Figure~\ref{fig:wispit2_compare_clean_nn}, middle panel).
To form a restored image, we supply the SIFARI model to CASA \texttt{tclean} as a \texttt{startmodel} with \texttt{niter=0}.
This convolves the model with the restoring beam and adds the residual image from the self-calibrated visibilities without further CLEAN iterations (Figure~\ref{fig:wispit2_compare_clean_nn}, right panel).

Both the SIFARI model and the restored SIFARI image show fewer spatial-filtering artifacts than the CLEAN image made before self-calibration.
The SIFARI model recovers a total flux of 190 mJy, closer to the flux measured in recent ALMA observations incorporating shorter baselines \citep[$210\pm21$~mJy;][]{benisty2026mapping}.
This closer agreement supports improved recovery of the ring's integrated emission despite the limited short-baseline coverage.

The RMS noise in the restored, self-calibrated SIFARI image (right panel) is roughly 30\% lower than in the original CLEAN image (left panel).
This application demonstrates that a SIFARI reconstruction can supply the sky model for phase self-calibration, extending the method's role to the calibration of the observations themselves.

\section{Discussion}\label{sec:discussion}
In the tests presented here, SIFARI resolves compact structure and recovers extended emission more faithfully than CLEAN.
The point-source and two-source tests demonstrate its ability to represent fine structure, while the uniform-disk tests show improved flux recovery when short baselines are missing (Sections~\ref{sec:psf}--\ref{sec:spatial_filtering_flux_recovery}).
The morphology benchmarks and PDS~70 application extend this comparison to complex emission and faint compact features near bright structure.
The WISPIT~2 application demonstrates a further use of the reconstruction: supplying a sky model for phase self-calibration when the CLEAN model is inadequate.

The coordinate-network representation reduces the number of quantities that must be fitted to describe a finely sampled image.
In pixel-based RML, each pixel intensity is a fitted parameter, so a large field sampled at high angular resolution can require millions of parameters.
SIFARI instead fits network parameters shared across sky positions, whose number is independent of the grid used to evaluate the image.
The architecture acts as an implicit morphological prior, guiding the recovery of emission on spatial scales that are weakly constrained by the measured visibilities.
The same architecture represents point sources, rings, and diffuse clouds in our tests without explicit spatial penalty terms such as Total Variation or Maximum Entropy.

A small Fourier feature scale can suppress compact emission, whereas a large scale can introduce high-frequency artifacts.
Our visibility-based estimator selects this scale without manual adjustment for each source, using an empirical mapping from the radial visibility-SNR profile.
Applying the same rule to each dataset makes scale selection reproducible and removes a source-specific tuning step.
This could help incorporate SIFARI into automated imaging pipelines.

The current estimator nevertheless remains an empirical prescription.
The comparisons in Appendix~\ref{app:auto_sigma} (Figure~\ref{fig:compare_sigma}) show that the automatic choice does not consistently maximize reconstruction scores, and that higher global scores do not necessarily imply clearer recovery of local features.
Establishing how broadly the rule can be used requires testing across a wider range of source morphologies, $uv$ coverages, and signal-to-noise levels.

The chosen Fourier feature scale does not describe directional differences in $uv$ coverage, because the estimator uses a radial visibility-SNR profile and the feature directions are sampled isotropically.
A single isotropic scale may therefore be inadequate for strongly anisotropic observations.
A possible extension is to select Fourier frequencies from the two-dimensional visibility distribution, allowing feature scales to vary by direction.

Alongside the best-fitting image, SWAG provides a spatially resolved measure of brightness variation among sampled reconstructions, which we use as an approximate uncertainty estimate.
Together with a thermal-noise floor, this estimate defines the SNR maps used to examine extended and compact emission.
The PDS~70 application illustrates how these maps supplement a background RMS estimate with information about the variability of reconstructed emission at each position.
Further validation should test how well these uncertainty estimates track reconstruction errors under different observing conditions.

Our findings indicate that machine-learning-driven image reconstruction has the potential to transform interferometric imaging, although several components of SIFARI still require refinement. 
The treatment of point-like sources is still largely empirical: we rely on the point-source response and two-source separation experiments to inform the grid selection for the other synthetic benchmarks (Sections~\ref{sec:psf} and \ref{sec:angular_resolution}). 
A clearer characterization of the effective response is necessary to anticipate its width without running a dedicated point-source test. 
Connecting this response to the intrinsic source morphology and the $uv$ coverage would enable the choice of a grid that adequately samples the reconstructed structure while avoiding unnecessary computational expense.

Training duration also remains an empirical choice.
We use a fixed Stage 1 budget and select the model with the lowest full-data loss, while checking that the loss and reconstructed images stabilize (Section~\ref{sec:training}).
The agreement with the known synthetic images supports this procedure for the cases studied here.
More robust convergence diagnostics would help determine the training needed for more complex fields.

Although the parameter count is independent of the image grid, larger grids still increase training computational cost.
On a single NVIDIA A100 GPU, the multiple-ring benchmark with a $790\times790$ reconstruction grid takes approximately 30~min for Stage 1 training (300 epochs; Section~\ref{sec:training}) and 50~min for the Stage 2 SWAG phase (500 epochs; Section~\ref{sec:uncertainty_quantification}), totaling about 80~min.
The inclined-ring and Ophiuchus benchmarks use smaller reconstruction grids and take less time.
Preliminary tests not included in this paper suggest that reconstructing morphologies more complex than the L1688 cloud model requires larger networks.
Available GPU memory also limits the grid size and network capacity.

The current implementation of SIFARI is restricted to 2D, single-pointing, single-polarization continuum observations.
Its forward model does not include mosaic observations or wide-field effects.
Extending the method to those observations would require the corresponding measurement effects to be included in the forward model.
Spectral-line cubes and polarization offer another direction: a joint reconstruction would require a sky model that accounts for frequency dependence or includes multiple Stokes components.

Finally, it is worth mentioning that fitting visibilities directly also provides a route to interferometric applications beyond the ALMA data studied here.
For observations that rely on closure quantities, such as those from the Event Horizon Telescope or JWST NIRISS aperture-masking interferometry, the forward model and loss would need to predict and compare these quantities.
Such an extension could retain the continuous sky representation while adapting the fit to the measurements available from each instrument.

\section{Conclusion}
\label{sec:conclusion}
SIFARI models sky brightness as a continuous coordinate-based function, fits measured visibilities without an external image training set, and sets the Fourier feature scale from the visibility data before training.
Sampling network weights with SWAG provides approximate pixel-wise brightness uncertainty estimates, which we combine with the thermal-noise term to construct SNR maps.
Synthetic ALMA benchmarks and real observations demonstrate the following results:
\begin{itemize}
\item {\it Angular Resolution:} In synthetic observations covering baselines between 15 m and 13.9 km, SIFARI produces an effective point-source response with a FWHM of $0.0040\arcsec \times 0.0037\arcsec$, about 8$\times$ and 4$\times$ narrower than the natural- and uniform-weighting CLEAN restoring beams, respectively.
It recovers two peaks at separations of $0.009\arcsec$ and $0.010\arcsec$, with the 50\%-midpoint criterion placing the effective separation scale between these values (Section~\ref{sec:angular_resolution}).
\item {\it Flux Recovery and Image Fidelity:} Under strong spatial filtering, SIFARI recovers 202 and 219 mJy from 200 mJy uniform disks with diameters of $0.6\arcsec$ and $1.0\arcsec$, respectively, compared with 41 and 20 mJy from CLEAN.
For the three morphology benchmarks, SIFARI achieves higher IFS and SSIM than the restored CLEAN images.
\item {\it High Dynamic Range Imaging (PDS~70):} The SIFARI reconstruction has a cleaner background inside the ring than the CLEAN image, making compact emission at PDS~70~b and c easier to distinguish.
The SWAG-based maps give SNR values above 5 for compact emission at the locations of PDS~70~b and PDS~70~c, and an average SNR of 23 for the bright outer ring.
\item {\it Self-Calibration Support (WISPIT~2):} SIFARI provides a self-calibration model for this long-baseline dataset, for which the CLEAN reconstruction does not provide a suitable model.
The restored, self-calibrated SIFARI image has fewer artifacts and an RMS noise approximately 30\% lower than the CLEAN image made from the original visibilities without self-calibration.
\end{itemize}
The visibility-based estimator provides a reproducible way to set the Fourier feature scale, although its empirical prescription does not consistently maximize reconstruction scores.
The tests support the use of neural visibility fitting to recover emission under strong spatial filtering, distinguish compact emission near bright extended structures, and construct models for self-calibration.

\section*{Code Availability}
The source code used in this work will be publicly released on GitHub and archived on Zenodo upon publication of this article.
During peer review, the code can be made available confidentially to the editor and referees upon request.

\begin{acknowledgments}
S.M. and A.I. acknowledge support from the National Aeronautics and Space Administration under grant No. 80NSSC18K0828.
Research presented in this article was supported by the Laboratory Directed Research and Development program of Los Alamos National Laboratory under project number 20250553ER. This paper is published under LA-UR-26-28006. This research is partially supported by the
National Nuclear Security Administration (NNSA) Advanced Simulation and Computing (ASC) Program under grant number DE-SCL0000089.
This research was enabled in part by support provided by the National Energy Research Scientific Computing Center (NERSC) under award number FES-ERCAP-m4239 and ERCAP0035555. 
The National Radio Astronomy Observatory and Green Bank Observatory are facilities of the U.S. National Science Foundation operated under cooperative agreement by Associated Universities, Inc.
We thank Myriam Benisty for providing the calibrated PDS~70 data, Kwang Moo Yi for insightful discussions on AI methods and ideas, and Ka Wai Ho for making computational resources available for this work. We also thank Stefano Facchini, Jessica Speedie, Jaehan Bae, Christopher M.~Johns-Krull, Charlie Gardner, Ryan Loomis, Eric Murphy, Brian Mason, Brian Kirk, Sifan Wang, Tianao Li, and Ruobing Dong for valuable discussions.

OpenAI Codex was used to assist with language editing of the manuscript. The authors take responsibility for the final content.
\end{acknowledgments}

\facilities{ALMA}

\software{PyTorch \citep{ansel2024pytorch},
          CLEAN and Multiscale CLEAN \citep{hogbom1974aperture,cornwell2008multiscale}
}

\clearpage

\appendix
\onecolumnfootnotes

\section{A Heuristic Estimator for the Fourier Feature Scale}\label{app:auto_sigma}

We set the Fourier feature scale from a radial visibility-SNR profile.
The estimator smooths this profile, identifies a characteristic support radius, and converts that radius to the half-normal scale used in Section~\ref{sec:fourier_feature_mapping}.

We group visibilities into 40 logarithmic bins spanning the observed range of positive $uv$ radii and evaluate the profile at the geometric bin centers, with radius $r$ expressed in arcsec$^{-1}$.
For each bin, the weighted signal power is
\begin{equation}
    P_\mathrm{sig}(r) =
    \left< |V|^2 - \frac{2}{W} \right>_W,
\end{equation}
where $\langle\cdot\rangle_W$ denotes averaging with the visibility weights $W$, and the subtraction removes the expected real-plus-imaginary noise power.
The corresponding per-visibility SNR is
\begin{equation}
    \mathrm{SNR}_\mathrm{vis}(r) =
    \frac{\sqrt{\max(P_\mathrm{sig}(r),0)}}{\sigma_\mathrm{noise}(r)}.
\end{equation}
For a bin containing $n$ visibilities, we use $\sigma_\mathrm{noise}(r)=n/\sum_{l\in\mathrm{bin}}\sqrt{W_l}$.
This profile measures SNR per visibility rather than the detection significance of the whole bin.

We smooth the profile to reduce fluctuations between neighboring radial bins.
Bins with at least $\max(10,\lceil0.001N_\mathrm{vis}\rceil)$ visibilities contribute to a centered, five-bin moving average of this profile.
The smoothing weight for each eligible bin is $\sqrt{n/\mathrm{median}(n)}$, clipped to $[0.5,2]$, where the median is taken over eligible bins.

To avoid treating a local SNR minimum as the end of the signal support, we form an envelope equal to the largest smoothed SNR at each radius or any larger radius.
The first crossing of this envelope below SNR=1 gives an initial estimate of the signal support radius $r_\mathrm{signal}$.

For bright sources, the SNR can remain above unity even after declining substantially from its low-$uv$ value.
We therefore also limit $r_\mathrm{signal}$ using a relative decrease in SNR.
We define the low-$uv$ SNR as the median smoothed value in the first three eligible bins, and cap $r_\mathrm{signal}$ at the first radius where the smoothed profile falls below 25\% of this value.
Where the bracketing SNR values are positive, we interpolate threshold crossings in log radius and log SNR.

We further limit the support radius to the sampled $uv$ coverage.
For the datasets studied here, the final support radius is
\begin{equation}
    r_\mathrm{support} =
    \min(r_\mathrm{signal}, r_{\mathrm{uv},99}),
\end{equation}
where \(r_{\mathrm{uv},99}\) is the 99th percentile of observed $uv$ radius.\footnote{The implementation also caps $r_\mathrm{support}$ at $1/(4\Delta x)$, where $\Delta x$ is the reconstruction pixel size; this safeguard is inactive in the experiments reported here.}
We convert this radius to the half-normal Fourier feature scale using the empirical relation
\begin{equation}
    \sigma_\nu = \frac{r_\mathrm{support}}{2}.
\end{equation}

\begin{figure*}[t!]
    \centering
    \begin{minipage}{0.95\textwidth}
        \centering
        \includegraphics[width=\textwidth]{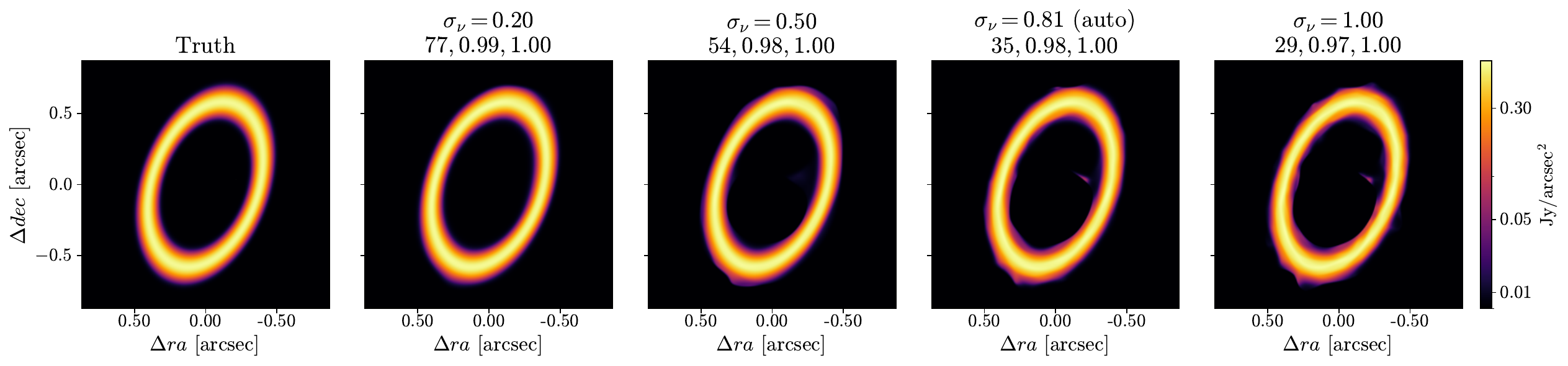}
        \\
        \includegraphics[width=\textwidth]{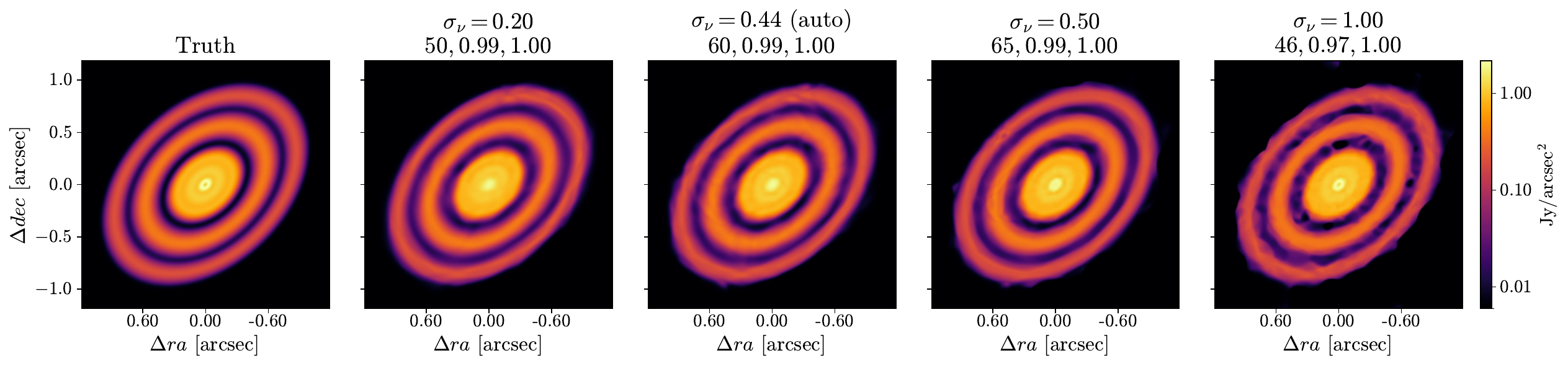}
        \\
        \includegraphics[width=\textwidth]{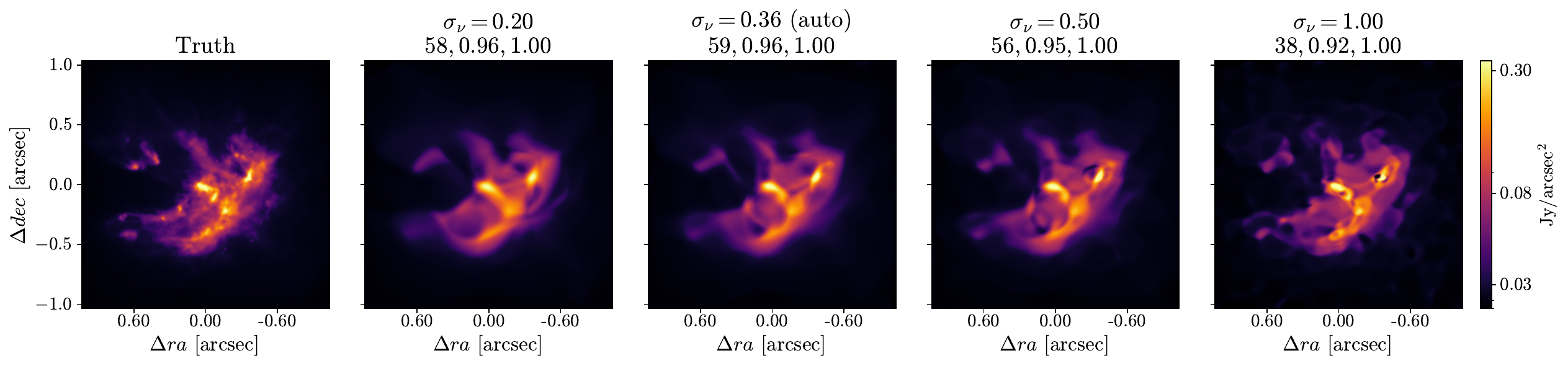}
    \end{minipage}
    \caption{
        Comparison of SIFARI reconstructions for different Fourier feature scales.
        Rows (top to bottom) show the inclined-ring, multiple-ring, and molecular-cloud benchmarks, with a common logarithmic intensity scale within each row.
        The leftmost column shows the ground truth; the remaining panels show reconstructions at the Fourier feature scales \(\sigma_\nu\) indicated in their titles. The scale selected by the visibility-supported heuristic estimator is marked ``(auto).''
        Panel-header annotations report image fidelity (IFS; left), SSIM (center), and recovered flux ratio (right).
    }
    \label{fig:compare_sigma}
\end{figure*}

Figure~\ref{fig:compare_sigma} compares reconstructions obtained with the automatic Fourier feature scale and fixed scales for the three morphology benchmarks.
The automatic scale yields the highest or near-highest IFS among the tested scales for the multiple-ring and molecular-cloud benchmarks.
For the inclined ring, smaller fixed scales yield higher IFS, but the injected point source is more clearly recovered with the automatic scale than with \(\sigma_\nu=0.50\).
Thus, the automatic choice does not consistently maximize reconstruction scores, and a higher global score does not necessarily imply clearer recovery of a local feature.

\bibliography{main}{}
\bibliographystyle{aasjournalv7}

\end{document}